\documentclass[12pt,preprint]{aastex}

\shorttitle{High Energy Observations of XRF~030723}
\shortauthors{Butler et al.}

\begin{document}

\title{High Energy Observations of XRF~030723: Evidence for an Off-axis Gamma-Ray Burst?}

\author{
N. R. Butler\altaffilmark{1},
T. Sakamoto\altaffilmark{3,4},
M. Suzuki\altaffilmark{3},
N. Kawai\altaffilmark{3,4},
D. Q. Lamb\altaffilmark{2},
C. Graziani\altaffilmark{2},
T. Q. Donaghy\altaffilmark{2},
A. Dullighan\altaffilmark{1},
R. Vanderspek\altaffilmark{1},
G. B. Crew\altaffilmark{1},
P. Ford\altaffilmark{1},
G. Ricker\altaffilmark{1},
J-L. Atteia\altaffilmark{6},
A. Yoshida\altaffilmark{4,7},
Y. Shirasaki\altaffilmark{8},
T. Tamagawa\altaffilmark{4},
K. Torii\altaffilmark{17},
M. Matsuoka\altaffilmark{9},
E. E. Fenimore\altaffilmark{5},
M. Galassi\altaffilmark{5},
J. Doty\altaffilmark{1},
J. Villasenor\altaffilmark{1},
G. Prigozhin\altaffilmark{1},
J. G. Jernigan\altaffilmark{10},
C. Barraud\altaffilmark{6},
M. Boer\altaffilmark{11},
J-P. Dezalay\altaffilmark{11},
J-F. Olive\altaffilmark{11},
K. Hurley\altaffilmark{10},
A. Levine\altaffilmark{1},
F. Martel\altaffilmark{1},
E. Morgan\altaffilmark{1},
S. E. Woosley\altaffilmark{12},
T. Cline\altaffilmark{13},
J. Braga\altaffilmark{14},
R. Manchanda\altaffilmark{15},
and G. Pizzichini\altaffilmark{16}}

\altaffiltext{1}{Center for Space Research, Massachusetts Institute
of Technology, 70 Vassar Street, Cambridge, MA, 02139}
\altaffiltext{2}{Department of Astronomy and Astrophysics, University
of Chicago, IL, 60637}
\altaffiltext{3}{Department of Physics, Tokyo Institute of Technology,
2-12-1 Ookayama, Meguro-ku, Tokyo 152-8551, Japan}
\altaffiltext{4}{RIKEN (Institute of Physical and Chemical Research),
2-1 Hirosawa, Wako, Saitama 351-0198, Japan}
\altaffiltext{5}{Los Alamos National Laboratory, P.O. Box 1663, Los
Alamos, NM, 87545}
\altaffiltext{6}{Laboratoire d'Astrophysique, Observatoire
Midi-Pyre\'ne\'es, 14 Ave. E. Belin, 31400 Toulouse, France}
\altaffiltext{7}{Department of Physics, Aoyama Gakuin University,
Chitosedai 6-16-1, Setagaya-ku, Tokyo 157-8572, Japan}
\altaffiltext{8}{National Astronomical Observatory, Osawa 2-21-1,
Mitaka, Tokyo 181-8588, Japan}
\altaffiltext{9}{Tsukuba Space Center, National Space Development
Agency of Japan, Tsukuba, Ibaraki, 305-8505, Japan}
\altaffiltext{10}{University of California at Berkeley, Space Sciences
Laboratory, Berkeley, CA, 94720-7450}
\altaffiltext{11}{Centre d'Etude Spatiale des Rayonnements, CNRS/UPS,
B.P.4346, 31028 Toulouse Cedex 4, France}
\altaffiltext{12}{Department of Astronomy and Astrophysics, University
of California at Santa Cruz, 477 Clark Kerr Hall, Santa Curz, CA 95064}
\altaffiltext{13}{NASA Goddard Space Flight Center, Greenbelt, MD, 20771}
\altaffiltext{14}{Instituto Nacional de Pesquisas Espaciais, Avenida
Dos Astronautas 1758, Sa\~o Jose\' dos Campos 12227-010, Brazil}
\altaffiltext{15}{Department of Astronomy and Astrophysics, Tata
Institute of Fundamental Research, Homi Bhabha Road, Mumbai, 400 005, India}
\altaffiltext{16}{Consiglio Nazionale delle Ricerche (IASF), via Piero
Gobetti, 101-40129 Bologna, Italy}
\altaffiltext{17}{Department of Earth and Space Science,
Graduate School of Science, Osaka University
1-1 Machikaneyama-cho, Toyonaka, Osaka 560-0043, Japan}

\def\eg{{\it e.g.}}
\newcommand{\ie}{{\it i.e.\,}}
\def\etal{{\it et al.\,}}
\def\et{{\it et al.\,}}
\def\etc{{\it etc.}}
\def\gcn{GCN}
\def\gtrsim{\mathrel{\hbox{\rlap{\hbox{\lower4pt\hbox{$\sim$}}}\hbox{$>$}}}}
\def\lessim{\mathrel{\hbox{\rlap{\hbox{\lower4pt\hbox{$\sim$}}}\hbox{$<$}}}}
\def\arcmin{$^{\prime}$}
\def\arcsec{$^{\prime\prime}$}
\def\degree{$^{\circ}$}
\def\Vsection{\vspace{-0.11in}\section}
\def\Vsubsection{\vspace{-0.11in}\subsection}
\def\Vsubsubsection{\vspace{-0.11in}\subsubsection}
\newcommand\hete{{\it HETE-2}}
\newcommand\xmm{{\it XMM-Newton}}
\newcommand\chandra{{\it Chandra}}
\newcommand\beppo{{\it BeppoSAX}}

\begin{abstract}
We report {\it High Energy Transient Explorer 2 (HETE-2)}
Wide Field X-ray Monitor/French Gamma Telescope observations 
of XRF~030723 along with observations of the XRF afterglow 
made using the 6.5m Magellan Clay telescope and the
{\it Chandra X-ray Observatory}.  The observed peak energy
$E^{\rm obs}_{\rm pk}$ of the $\nu F_{\nu}$ burst spectrum is 
found to lie
within (or below) the WXM 2-25 keV passband at 98.5\% confidence, 
and no counts are detected above 30 keV.  Our best fit value
is $E^{\rm obs}_{\rm pk}=8.4^{+3.5}_{-3.4}$ keV. The ratio of X-ray
to $\gamma$-ray flux for the burst follows a correlation found
for GRBs observed with \hete, and the duration of the burst is
similar to that typical of long-duration GRBs.  If we require
that the burst isotropic equivalent energy $E_{\rm iso}$ and 
$E_{\rm pk}$
satisfy the relation discovered by \citet{amati02}, 
a redshift of $z=0.38^{+0.36}_{-0.18}$ 
can be determined, in agreement with constraints
determined from optical observations.  We are able to fit the
X-ray afterglow spectrum and to measure its temporal fade.  
Although the best-fit fade is shallower than
the concurrent fade in the optical, the spectral similarity
between the two bands indicates that the X-ray fade may actually 
trace the optical fade.  If this is the case, the late
time rebrightening observed in the optical cannot be due to a 
supernova bump.  We interpret the prompt and afterglow X-ray
emission as arising from a jetted GRB observed off-axis and possibly
viewed through a complex circumburst medium due to a progenitor
wind.
\end{abstract}


\section{Introduction}

Approximately 1/3 of all bursts detected by the \hete~satellite,
excluding X-ray bursts and bursts due to known Galactic sources, 
have fluences in the X-rays (2-30 keV) which are larger than those
measured in the $\gamma$-rays (30-400 keV).  These are the
``X-ray Flashes'' (XRFs) \citep{heise2000}.  A similar
fraction of XRFs to $\gamma$-ray Bursts (GRBs) was observed by
{\it GINGA} \citep{strohmayer1998}, while {\it BATSE} lacked the low energy
response to detect so large a fraction of XRFs relative to GRBs.
Nonetheless, for bursts detected by both {\it BATSE} and {\it BeppoSAX},
it appears that XRFs can be described by \citet{band93} models with 
parameters consistent with those
found for GRBs, except with lower fluxes and $\nu F_{\nu}$ peak 
energies $\sim$10--100 times smaller \citep{kippen2002}.

\citet{sakamoto03} have demonstrated additional links between
XRFs and GRBs by showing that the prompt emission from XRF~020903,
the first XRF with an established redshift \citep{soder04a}, satisfies
and extends two correlations observed for GRBs.  First, the burst
satisfies the relation
found by \citet{barraud03} between 7-30 keV and 30-400 keV fluences
in bursts observed by the \hete~French Gamma Telescope (FREGATE).
Second, the isotropic-equivalent energy $E_{\rm iso}$ tracks the
$\nu F_{\nu}$ spectral peak energy $E_{\rm pk}$ as found for GRBs
by \citet{amati02}, extending the correlation by a factor $\sim$300.

Much theoretical (and observational) energy has focused recently on
XRFs, driven by the strong possibility that XRFs represent an extension
of the GRB population to softer energies.  Models are focused
primarily on differences in the jet physics
(see \citet{zhang2002} for an overview).  
The data for XRF~030723, and the well-sampled data 
from its optical and X-ray afterglows, are excellent discriminators
for these models, as we discuss in Sections \ref{sec:opt_aft},
\ref{sec:030723_dark}, and \ref{sec:discuss}.

\section{Prompt Observations}

\subsection{Localization}

\begin{figure}[ht]
\centering
\resizebox{30pc}{!}{\includegraphics{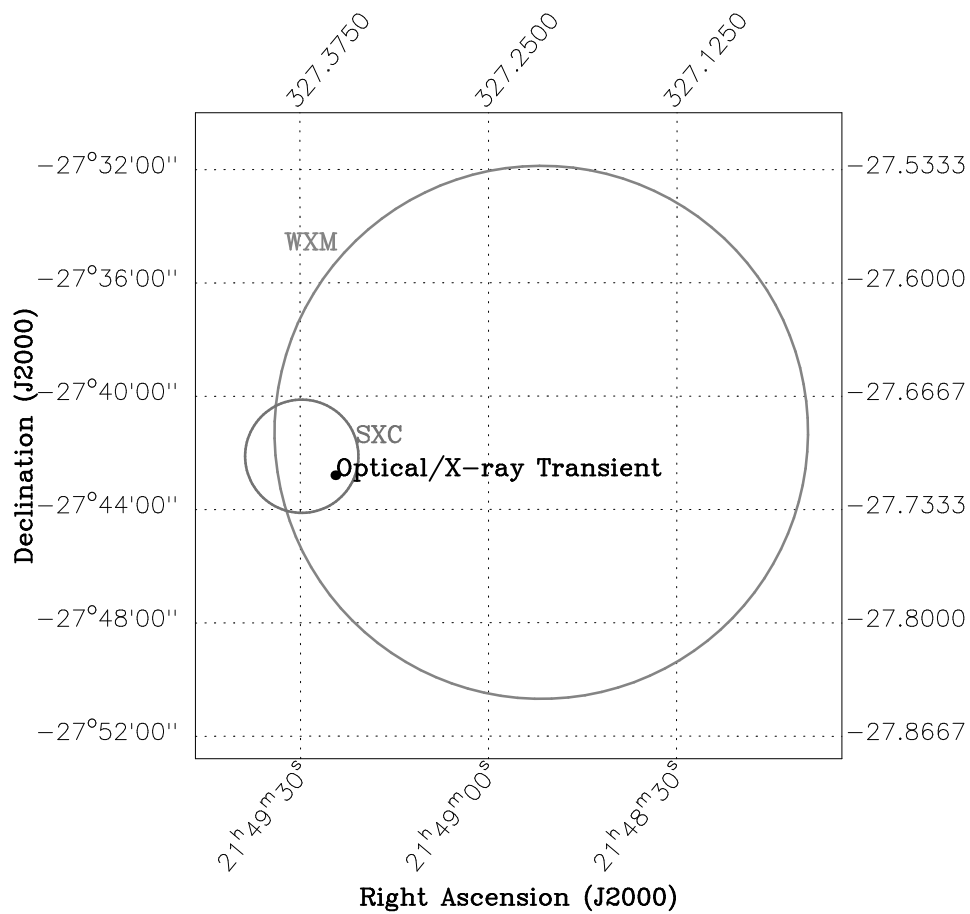}}
\caption{
\small
\hete~WXM/SXC localization for XRF~030723.  The larger circle is
the 90\% confidence WXM error region, and the smaller circle is the
SXC 90\% confidence error region \citep{prig03}.  The final 
\hete~error region is
the SXC error region.  The point labeled
``Optical/X-ray Transient'' refers to the location of the spatially 
coincident optical and
X-ray afterglows to XRF~030723 reported by \citet{fox03} and by
\citet{butler03b}, respectively. 
}
\label{fig:skymap}
\end{figure}

XRF~030723 was detected by the \hete~FREGATE, WXM (Wide Field
X-ray Monitor), and SXC (Soft X-ray Camera)
instruments at 6:28:59 UT on 2003 July 23 \citep{prig03}.
A GCN burst alert was issued 42s later, reporting a flight-derived 
WXM localization with a 30 arcmin radius (90\% confidence). Ground 
analysis of the WXM data produced a refined 9.4 arcmin error radius 
localization 
that was reported in a GCN Notice at 09:47:25 UT.
It is centered at R.A. $=21^h 48^m 52^s$, decl. $= -27^{\circ} 41' 16''$
(J2000.0).
Ground analysis of the SXC data provided an initial localization that 
was disseminated as a GCN Notice at 13:38:19 UT.  A refined 2 arcmin radius
error circle centered at R.A. 
$= 21^h 49^m 27.4^s$, decl. $= -27^{\circ} 42' 01''$  (J2000.0)
 was reported by \citet{prig03}.  These error regions are shown in
 Figure \ref{fig:skymap}.

In our analysis of the XRF~030723 prompt emission, 
we apply a ``cut'' to the WXM time- and energy-tagged 
data (TAG data), using only
the photons from the pixels on the five wires in the X-detector
(XA0, XA1, XA2, XB0, XB1) and the three wires in the Y-detector
(YA0, YA1, YA2) that were illuminated by the burst.  This entails including
only the photons from the portions of each wire which would be illuminated 
by an X-ray source located at the position of the optical transient (OT) 
reported by \citet{fox03} (Section \ref{sec:opt_aft}).

\subsection{Temporal Properties}

\begin{figure}[ht]
\centering
\includegraphics{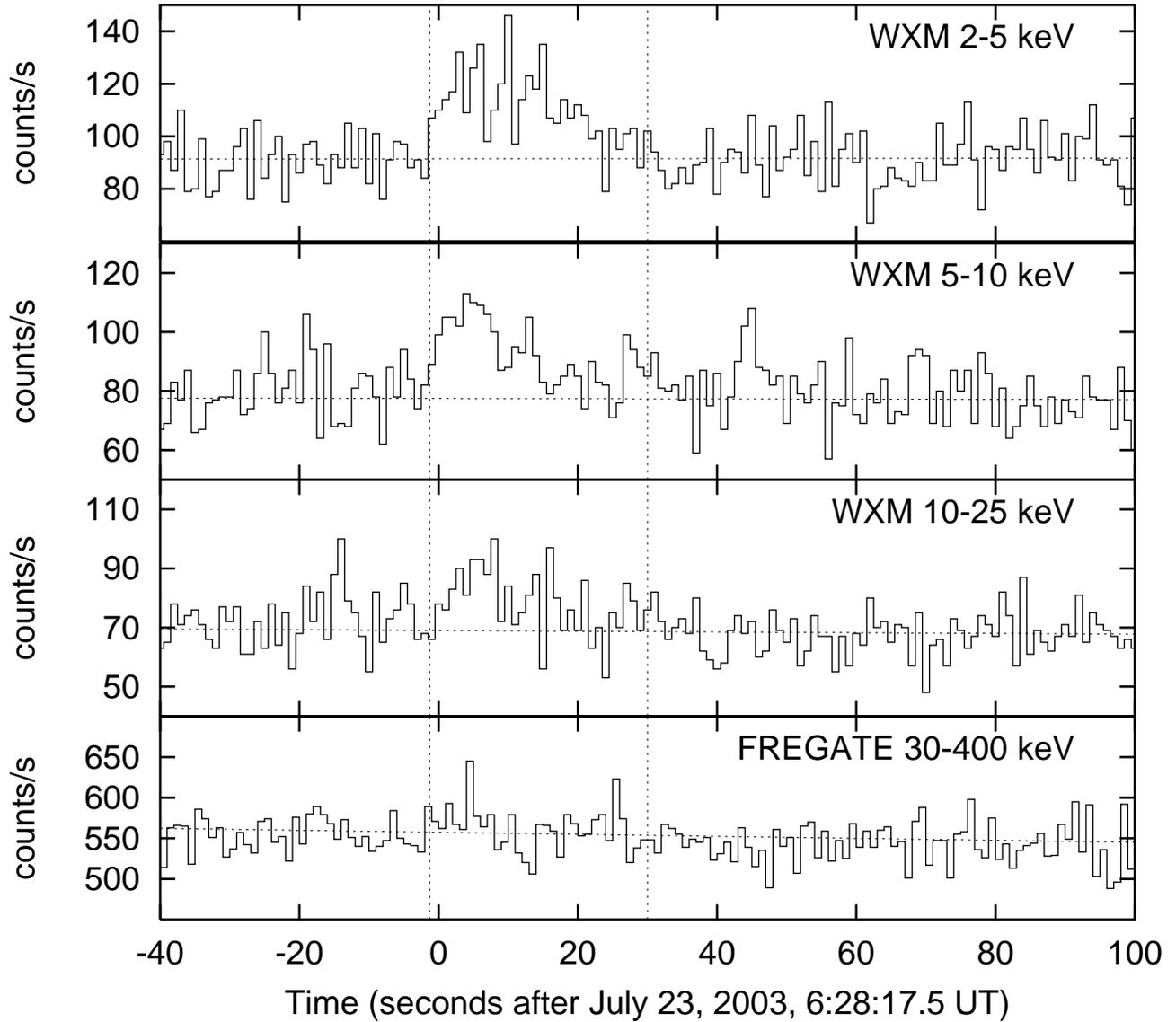}
\caption{
\small
Light curve for XRF~030723 in three WXM energy bands (2-5 keV, 5-10 keV,
10-25 keV) and for FREGATE in the 30-400 keV band.  The data are binned 
in one second intervals.
Source data for spectral fitting are extracted from the region between
the vertical dotted lines (between -1.25 and 30 seconds).  The background
fits are plotted as horizontal dotted lines.
}
\label{fig:wxm_lcs}
\end{figure}

Figure \ref{fig:wxm_lcs} displays the WXM and FREGATE light curves for 
XRF~030723 in
four energy bands.  A linear fit to the 30-400 keV data (4th panel, Figure
\ref{fig:wxm_lcs}) shows that there are few, if any, net counts above 30 keV.
The burst temporal profile below 25 keV is consistent with the 
presence of a single peak, and the burst durations in this band
do not appear to depend strongly on energy.  
Table 1 gives $t_{90}$ and 
$t_{50}$ durations for the burst in the 2-5, 5-10, 10-25, and 2-25 keV bands.

\subsection{Spectrum}
\label{sec:hete_spec}

The WXM detector response matrix has been well-calibrated using
observations of the Crab nebula \citep{shirasaki2002}.  In the spectral
fits, we include only the photons that registered on the five wires in
the X-detectors and the three wires in the Y-detectors that were
illuminated by the burst, as mentioned above.  Since the variation in
the gain is not uniform at the ends of the wires in the WXM detectors
\citep{shirasaki2000}, we use only the photon counts that registered in
the central $\pm$50 mm region of the wires to construct the spectra of
the burst.  We include all of the photons that register in the central
regions of these wires (i.e, we use the full 2-25 keV energy range of
the WXM).  We analyze the summed data from the eight wires, taking
the normalizations on all wires to be the same.  The FREGATE spectral
response has been verified with observations of the Crab nebula as
described in \citet{barraud03}.  We consider the
WXM data in the 2-25 keV band and the FREGATE data in the 7-100 keV
band, and we do not rebin the pulse height channels.
We carry out a set of fits for the burst from -1.25s to 
30s from the trigger time (see the dotted lines in Figure \ref{fig:wxm_lcs}),
that include the spectral data from both the WXM and the FREGATE.  
Background data are taken from the time intervals -37.2s to -14.9s and
44.2s to 99.4s, relative to the trigger time.  We use 
the XSPEC v11.2.0 software package to do the spectral fits.  All errors
refer to the 90\% confidence regions, unless otherwise noted.

\begin{figure}[ht]
\centering
\resizebox{30pc}{!}{\includegraphics{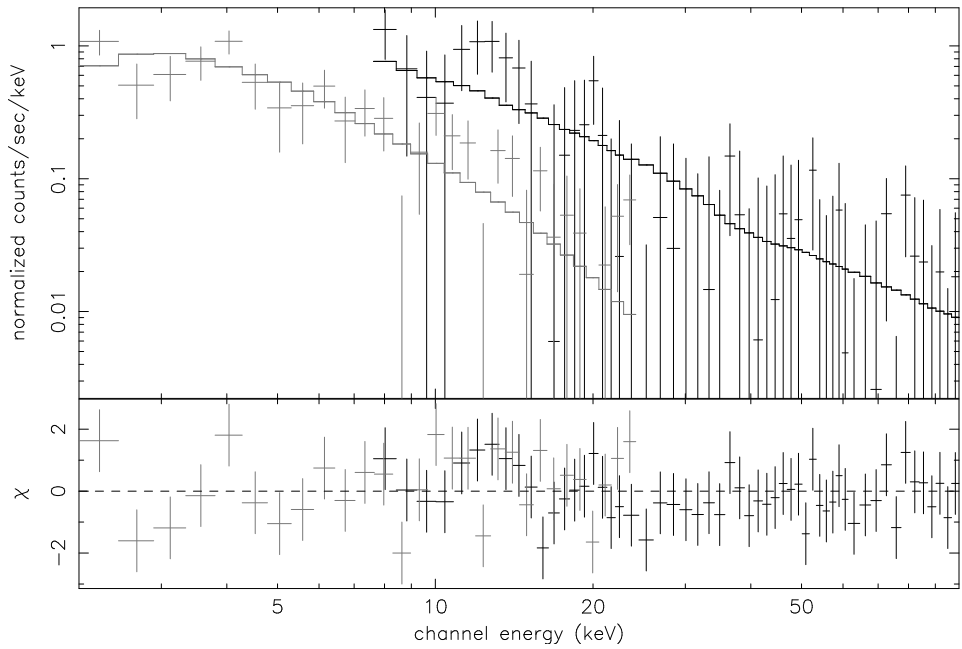}}
\caption{
\small
WXM and FREGATE spectra for XRF~030723.  Data are fitted jointly for
the WXM (2-25 keV) and FREGATE (7-100 keV) from the time interval -1.25 to 30 
seconds.  The model shown as histograms plotted with the data (crosses)
is the best-fit power-law model (Table \ref{table:spec_pars}).}
\label{fig:wxm_spec}
\end{figure}

Table \ref{table:spec_pars} presents the results of our time-averaged
spectral analysis for XRF~030723.  The data are well fitted by a
power-law model (Figure \ref{fig:wxm_spec}) and considerably less 
well-fitted by a blackbody model.  The data bins above 30 keV do
not affect the blackbody fit.  In the 2-30 keV band, the fit
yields $\chi^{2}_{\nu}=1.484$ for 50 degrees of freedom.  Thus, it is
rejectable at 98.5\% confidence.

Band models can be 
used to describe the spectra of essentially all GRBs.
The high-energy photon index of a Band model must
satisfy $\beta < -2$ in order for the $\nu F_{\nu}$ spectral peak energy to lie
in or below our passband.
The index in the best fit power-law model is consistent
with $\beta<-2$.  If we exclude 
the counts below 10 keV, retaining 66 of 84 data bins, the best-fit 
power-law index ($-3.2$) 
is less than $-2$ at 98.5\% confidence (likelihood ratio test: $\Delta 
\chi^2=5.93$ for 1 additional degree of freedom).  
This is the confidence at which we claim the burst to be an XRF. 
If the $\nu F_{\nu}$ peak energy of the burst were not established in 
this fashion to lie in or below the WXM band (2-25 keV), the burst 
could potentially be a faint GRB, with $E^{\rm obs}_{\rm pk}>100$  keV.

\begin{figure}
\centering
\resizebox{20pc}{!}{\includegraphics{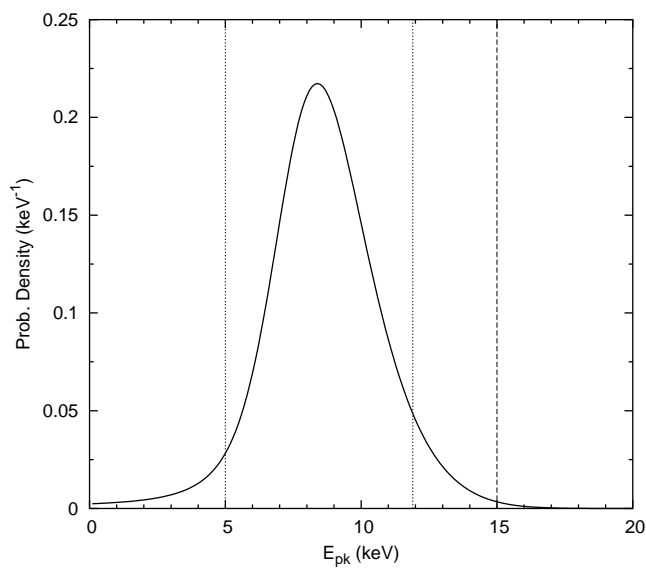}}
\caption{
\small
The posterior probability density distribution for $E^{\rm
obs}_{\rm pk}$ from the constrained Band model.  This is
calculated by numerically integrating $\exp{(-\chi^2/2)}$
over the constrained Band model normalization and $\beta$ parameters,
then normalizing the resulting $E^{\rm obs}_{\rm pk}$ distribution.
The vertical dotted lines define the 90\%
probability interval, and the vertical dashed line defines the
99.7\% probability upper
limit of 15 keV.  The peak of the probability occurs at 
$E^{\rm obs}_{\rm pk}= 8.4$ keV.
}
\label{fig:post_prob}
\end{figure}

A proper
determination of $E^{\rm obs}_{\rm pk}$ should be found using a
Band model.  The parameters of our best fit models are shown in
Table \ref{table:spec_pars}.  The 90\% confidence parameter regions
are not well defined for the Band models; so we quote the 68\% confidence
regions in Table \ref{table:spec_pars} instead.
If we assume $\beta<-2$, as argued above, we can generate firm limits
on $E^{\rm obs}_{\rm pk}$ using
the ``constrained'' Band model formalism developed by
\citet{sakamoto03}.  This model assumes $\beta<-2$ and allows
the Band model to produce a pure power-law only with the high
energy index.  We find $E^{\rm obs}_{\rm pk}=8.4^{+3.5}_{-3.4}$ keV.  
The posterior
probability distribution for $E^{\rm obs}_{\rm pk}$, along with the
locations of this
90\% confidence interval and the 3-$\sigma$ upper limit of 15.0 keV,
are plotted in Figure \ref{fig:post_prob}.
We find consistent values for $E^{\rm obs}_{\rm pk}$ using cutoff 
power-law model and Band model fit with $\alpha=-1$.  This is a typical 
$\alpha$ value for GRBs,
and it allows for excellent fits to the XRF~030723 data.

Table 3 displays fluences and peak fluxes 
determined from fits of the cutoff power-law model in various energy
bands.  The power-law model produces a lower value for the 30-400 keV
flux than does the cutoff power-law model.  Applying the best-fit 
power-law model in Table \ref{table:spec_pars}, the burst fluence in the 
2-30 keV band is $S_X({\rm 2-30 keV}) = 2.9 \pm 0.4
\times 10^{-7}$ erg cm$^{-2}$, and the burst fluence in the 30-400 keV band
is $S_{\gamma}({\rm 30-400 keV}) = 2.8 ^{+1.5}_{-1.3}
\times 10^{-7}$ erg cm$^{-2}$.  For both the cutoff power-law model and
the power-law model we have $\log{ [S_X({\rm 2-30 keV}) / 
S_{\gamma}({\rm 30-400 keV}) ]} > 0$, which is our operating definition of 
an XRF.  

Finally, we test for possible spectral evolution during the burst by
fitting the WXM and FREGATE data divided into two time portions.  We take
data from the first half of the burst (from -1.25s to 15s) and from the
second half of the burst (from 15s to 30s), and we fit both portions using
a power-law model in the same fashion as we fit the full data set above.  
A power-law fit to the first data half yields $\beta_1=-1.9^{+0.2}_{-0.1}$,
with $\chi^2_{\nu}=72.95/82$.  For the second half of the data, we find
$\beta_2=-2.9^{+0.8}_{-1.4}$, with $\chi^2_{\nu}=71.77/82$.  Fitting both
halves jointly and allowing one common spectral index, we find 
$\beta=\beta_1=\beta_2=-1.9^{+0.1}_{-0.2}$, with $\chi^2_{\nu}=150.3/165$.
Hence, there is evidence for a hard-to-soft spectral evolution during
the burst at 98.2\% confidence ($\Delta \chi^2 = 5.58$ for 1 additional 
degree of freedom).  Hard-to-soft evolution is common in GRBs, and it was 
reported by \citet{sakamoto03} at similar confidence for XRF~020903.

\section{Optical Afterglow Observations}
\label{sec:opt_aft}

The refined error 
regions are shown in Figure \ref{fig:skymap}, enclosing the location of a
fading OT  detected by \citet{fox03}.  The quoted position of the OT
is R.A. $=21^h 49^m 24.40^s$,  decl. $=-27^{\circ} 42' 47.4''$ (J2000.0),
with an uncertainty of less than 0.5'' in either dimension.
Using the 60-inch telescope at
Mount Palomar, \citet{fox03} observed the source to fade by 1.1 mag,
$R\sim 21.3$ 1.23 days after the GRB to $R\sim 22.4$ 2.23 days after the
GRB.   This source, likely the optical afterglow associated with the GRB,
was also detected by \citet{fox03} in the Ks-band ($Ks\sim 18.65$) with 
the Hale 200-inch
telescope at Mount Palomar 1.23 days after the burst, however the source
was below the detection threshold ($Ks\sim 19.0$) at 2.23 days.

From 24.8 hours to 25.2 hours after the burst (centered on July 24.31 UT),
we observed the SXC error circle with the LDSS2
instrument on the 6.5m Magellan Clay telescope at Las Campanas
Observatory in Chile.  Four 6-minute R-band exposures were
taken in $\sim 0.6$\arcsec~seeing.  
Our photometry has been calibrated against the USNO photometry data
reported by \citet{henden03}.
Coaddition of the images gives a
limiting magnitude of $R = 24.5$.  We detected the source of \citet{fox03}
with $R = 21.13 \pm 0.05$ \citep{dull03a}.
On July 28.385 UT, 5.13 days after the
burst, we again observed the SXC error circle with Magellan.
We obtained two 200-second exposures with the MagIC instrument in
$\sim 0.8$\arcsec~seeing, reaching a limiting magnitude of $R=24.3$.
In the second observation the OT magnitude was $R = 24.2 \pm 0.3$.
Including the other detections reported over the GCN (see Figure 
\ref{fig:lc_opt}), we estimate a late time power-law decay index of 
$\alpha=-1.84 \pm 0.04$, and an early power-law decay of 
$\alpha=-0.8 \pm 0.2$.  A break in the light curve can then be inferred 
to occur at $\sim$1.5 days
after the burst \citep[see also,][]{fynbo04a}.  
The simple two component power-law picture is, however,
inadequate to explain the early faintness (at $t\lesssim 0.5$ days) and 
the late time rebrightening in the optical afterglow.

\citet{rykoff03} argue that the early faintness in the optical
afterglow (relative to the afterglow after $\sim 1$ day) may be due to 
absorption by material arising from a progenitor wind.  However,
their measurements (left-most two optical points in Figure \ref{fig:lc_opt})
are of low significance, and they indicate an early-time flux {\it higher}
than that implied by the observations reported by \citet{fynbo04a} (next
two optical data points in Figure \ref{fig:lc_opt}).  \citet{fynbo04a} suggest
a complicated off-axis jet scenario that could possibly explain all four 
data points.

\citet{fynbo04a} \citep[see also,][]{fynbo04b,tom04}
observed a rebrightening in the R-band optical afterglow
at $t\gtrsim t_{\rm GRB}+9$ days.  Those authors argue
that this is due to emission from an underlying supernova component, 
and the magnitude of
the rebrightening implies that the supernova is nearby ($z\sim 0.6$).  
\citet{dado03} also argue in favor of this supernova component interpretation.
Alternatively, \citet{huang04}, and also \citet{liang04},
argue that the rebrightening is due to a two 
component jet as put forward
by \citet{berger03} for GRB~030329.  Regardless of how the rebrightening
is interpreted, a low
redshift would be consistent with the blue, apparently featureless VLT 
spectrum observed by \citet{fynbo04a}, where the lack of Ly-$\alpha$ forest
lines implies $z<2.3$.

\citet{fynbo04a} detected a faint ($R=26.8 \pm 0.4$) galaxy spatially
coincident with the optical afterglow using the
FORS1 optical camera on the ESO VLT on September 24 UT.  From September
29 UT to October 3 UT, \citet{kawai03} observed the probable host galaxy
with the Subaru Prime Focus Camera on the Subaru telescope and detected
the source in the Rc ($Rc=27.6 \pm 0.4$), Ic, and z' bands.  

The source was not detected in the radio.  \citet{soder03} report a
3-$\sigma$ upper limit of 180 $\mu$Jy at 8.46 GHz for July 26.42 UT.

\section{X-ray Afterglow Observations}
\label{sec:030723_dark}

\subsection{Chandra Detection}

On 25 July 2003, the {\it Chandra Observatory} targeted the field of
XRF~030723 for a 25 ksec (E1) observation spanning the
interval 09:52-17:05 UT on 25 July, 51.4 - 59.0 hours after the
burst. The SXC error circle from \citet{prig03} was completely
contained within the field-of-view of the \chandra~ACIS-I array.
On 4 August 2003, \chandra~re-targeted the field of
XRF~030723 for an 85 ksec followup (E2) observation, spanning the interval 4
August 22:22 UT to 5 August 22:27 UT, 12.69 to 13.67 days after the
burst.  For this observation, the SXC error circle from \citet{prig03}
was completely contained within the field-of-view of the \chandra~ACIS-S3 chip.

\begin{figure}[ht]
\centering
\resizebox{30pc}{!}{\includegraphics{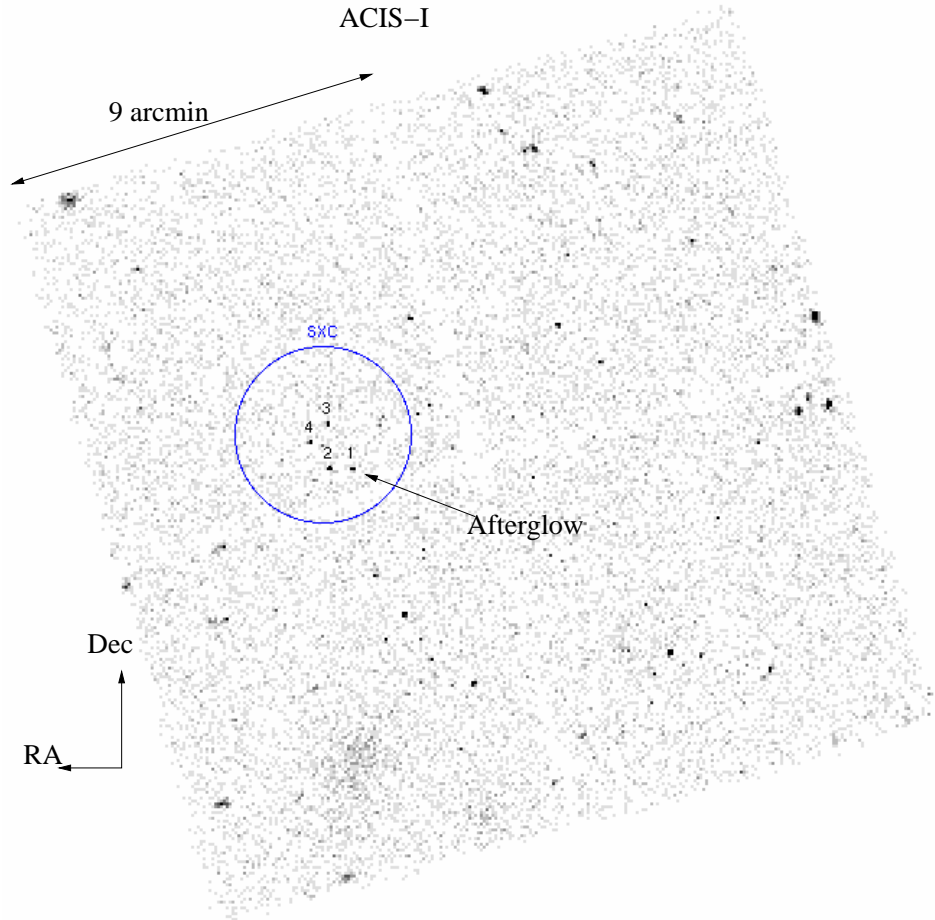}}
\caption{
\small
Three X-ray sources and one USNO field star (source \#2) were found within
the SXC error region in our E1 ACIS-I observation.  The brightest source 
\#1, spatially coincident with
a source later confirmed by \citet{fox03}, is labelled ``Afterglow.''
Our E2 observation with ACIS-S3 revealed that this source faded in
the X-ray's as well.
}
\label{fig:chandra_find}
\end{figure}

As reported in \citet{butler03a}, 3 candidate point sources
were detected within the revised SXC error region in our E1 observation
(Figure \ref{fig:chandra_find}).
Positions and other
data for these sources are shown in Table 4.
None of the sources
were anomalously bright relative to objects in \chandra~deep field
observations \citep[see, e.g.,][]{rosati02}.
The brightest object within the SXC error circle (source \#1),
lies 62\arcsec~from the center of the SXC error circle, and is within
0.7\arcsec~of the optical transient reported by \citet{fox03}.

Table 4 shows the number of counts detected in
E1 and in E2.  The E2 observations were reported in \citet{butler03b}.
Accounting for the difference in exposure times and sensitivity,
the number of counts detected for a steady source in E2 should be
$\sim 6$ times the number of counts detected in E1.  This factor
is greater than the ratio of exposure times, because the ACIS-S3 chip
is more sensitive than the ACIS-I chips.  (We estimate the additional
sensitivity by assuming a power-law spectrum with photon index $\Gamma=-2$ and
Galactic absorption.)
  Thus, sources
3 and 4 appear to have remained constant, while source 1 has faded.
The number of counts detected in E2 corresponds to a $\sim 7\sigma$
significance decrease (i.e.  factor of ~6) in flux since the E1 observation.

\subsection{X-ray Afterglow Spectrum and Fade}
\label{sec:xray_fit}

\begin{figure}[ht]
\centering
\resizebox{18pc}{!}{\includegraphics{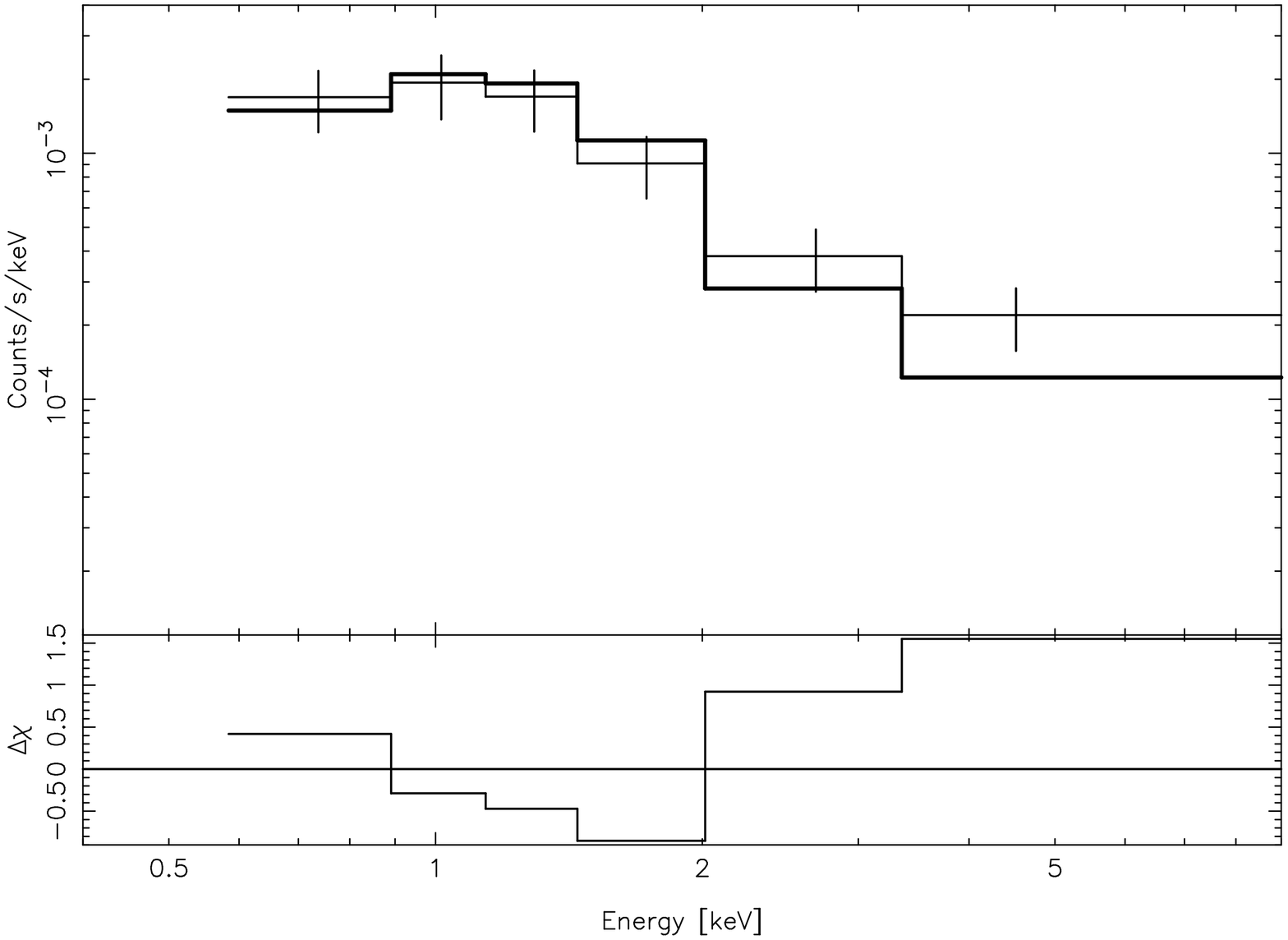}}
\resizebox{18pc}{!}{\includegraphics{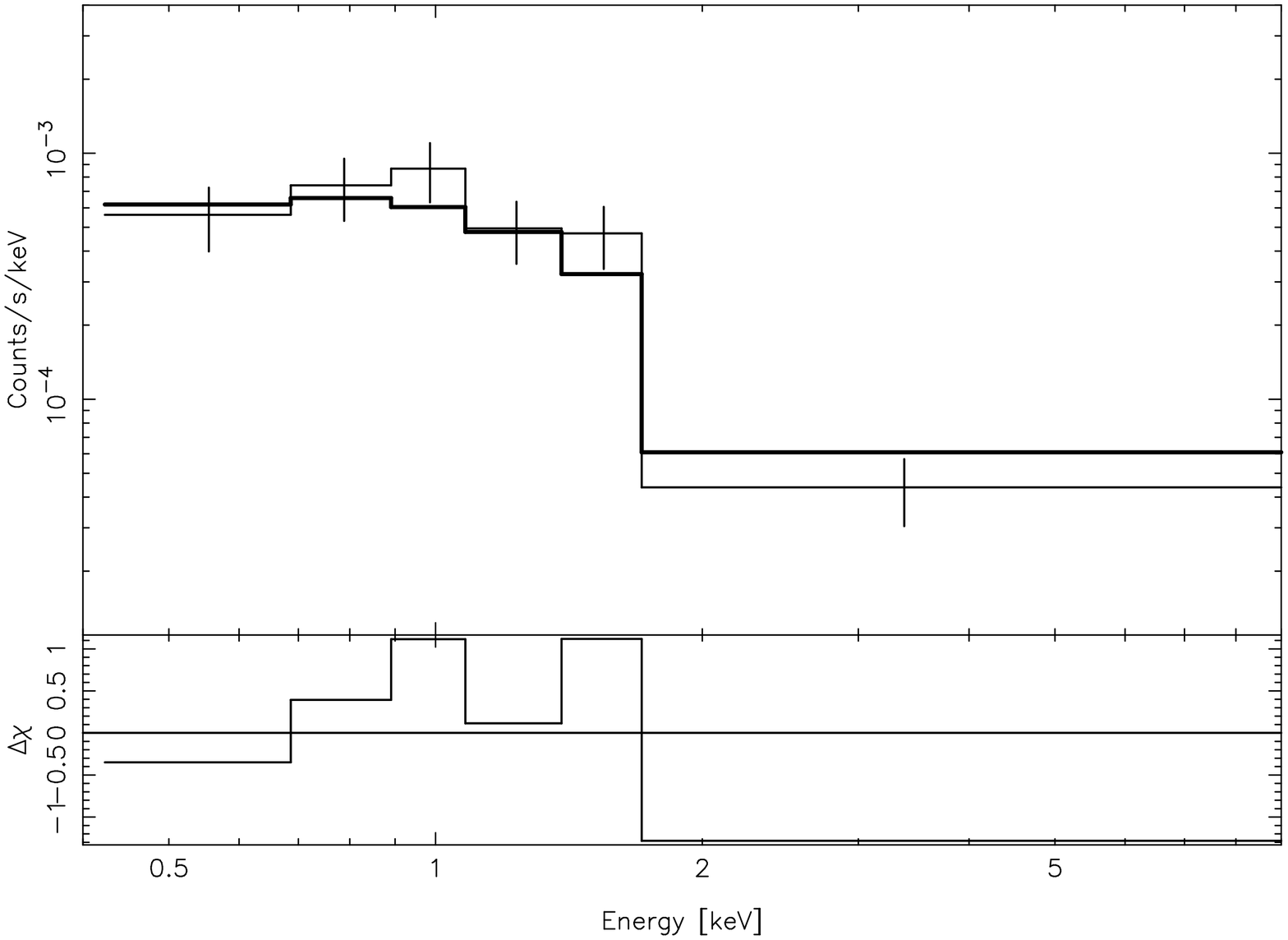}}
\caption{
\small
The counts in E1 (ACIS-I, left plot) and E2 (ACIS-S, right plot) are 
fitted simultaneously using an
absorbed power-law model ($\chi^2_{\nu} = 8.9/9$).
}
\label{fig:030723_sp}
\end{figure}

\begin{figure}[ht]
\centering
\rotatebox{0}{\resizebox{30pc}{!}{\includegraphics{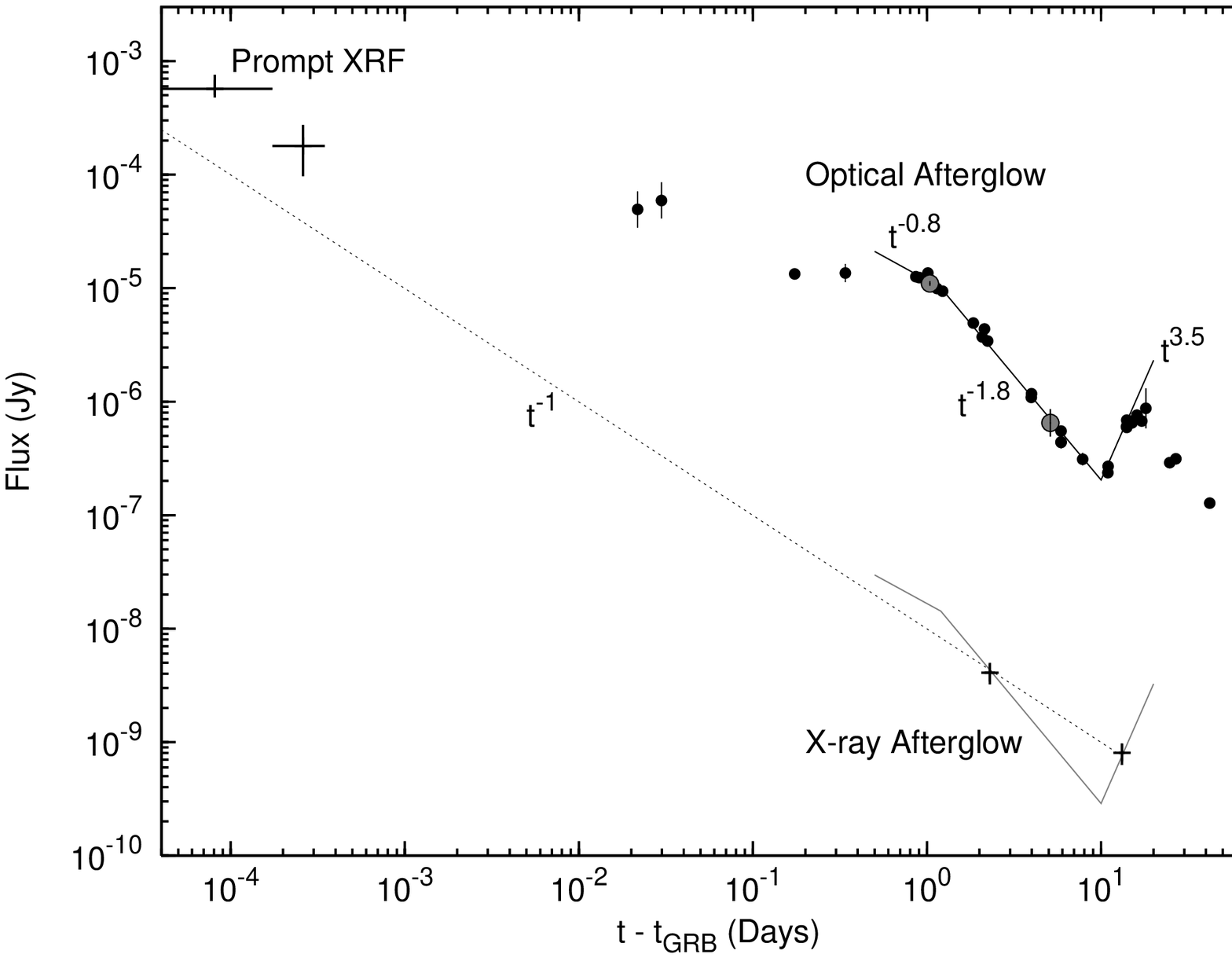}}}
\caption{
\small
The optical spectral flux and the X-ray spectral flux
(at 1 keV) for the afterglow to XRF~030723.  The optical data are taken 
from the literature
\citep{fox03,dull03a,dull03b,bond03,smith03,fynbo04a}.  Our data are
marked with open circles.  All optical data are from the R-band, except for
the left-most two points from \citep{smith03}, which are unfiltered.
We apply an extinction correction to the optical data of 0.088 mag
\citep{schleg98}.  A temporal break appears to be 
present in the
spectrum at $t\sim 1.5$ days \citep{dull03b}.  
A rebrightening is present after $t\sim 9$ days \citep{fynbo04a}.  We
overplot a 3-times broken power-law fit, to describe the optical behavior.  
The X-ray afterglow spectral flux may follow this 
same trend (gray line), or the behavior may be
that of a simple power-law fade (dotted line).  
}
\label{fig:lc_opt}
\end{figure}

To properly determine the fade factor for the X-ray afterglow,
we fit the E1 and E2 spectral
data jointly.  We reduce the spectral data using the standard
CIAO\footnote{http://cxc.harvard.edu/ciao/} processing tools.
We use
``contamarf''\footnote{http://space.mit.edu/CXC/analysis/ACIS\_Contam/script.htm
l} to correct for the quantum efficiency degradation due to
contamination in the ACIS chips, important for energies below $\sim 1$ keV.
We bin the data into 12 bins, each containing 12 or more counts,
and we fit an absorbed power-law model by minimizing $\chi^2$.  The
model has three parameters: two normalizations, and one photon index
$\Gamma$.  The absorbing column has been fixed at the Galactic value
in the source direction, $N_H = 2.4 \times 10^{20}$ cm$^{-2}$
\citep{dickey1990}.
The model fits the data well ($\chi^2_{\nu} = 8.9/9$,
Figure \ref{fig:030723_sp}).  The best fit photon number
index is $\Gamma = -1.9^{+0.2}_{-0.3}$, which is a typical
value for the X-ray afterglows of long duration GRBs \citep{costa99}.
Using this
model, we find that the E1 flux is $2.2 \pm 0.3 \times
10^{-14}$ erg cm$^{-2}$ s$^{-1}$ (0.5-8.0 keV band), while the E2 flux
is $3.5 \pm 0.5 \times 10^{-15}$ erg cm$^{-2}$ s$^{-1}$ (0.5-8.0 keV band).
The decrease in
flux between the two epochs can be described by a power-law with a
decay index of $\alpha= -1.0 \pm 0.1$.  This value of $\alpha$ is
consistent with the early power-law decline in the optical (Section
\ref{sec:opt_aft} for $t\lessim 1.5$ days after the GRB; however, the
index is considerably flatter than the index at $t>1.5$ days.
This flatter X-ray decay may possibly be related
to the rebrightening of the optical afterglow reported by \citet{fynbo04a}.
The optical afterglow, had it been observed during our two 
\chandra~epochs only, would indicate a fade with $\alpha\sim -1$
(see Figure \ref{fig:lc_opt}).

\section{Discussion}
\label{sec:discuss}

\subsection{Source Properties}

The WXM/FREGATE spectrum for XRF~030723 is poorly fit by a blackbody
model.  Even though the best fit model
had a temperature similar to that commonly found
in Type I X-ray burst sources \citep[see, e.g.,][]{lewin1993},
an association with an X-ray burster can be rejected due to the
source's high Galactic latitude ($b = -49.8^\circ$).  No known
X-ray burster or globular cluster is coincident with the position of 
the afterglow and its probable host galaxy.  On the contrary, the burst and 
afterglow energetics and 
fade properties are similar to those common in GRBs and XRFs, as we 
now discuss.

\citet{barraud03} find a strong correlation between the fluence in
the 7-30 keV versus the fluence in the 30-400 keV band for 32 GRBs
observed with HETE FREGATE.  \citet{sakamoto03}
find that XRF~020903 supports an extension of the correlation to XRFs.
The correlation is described as
$S_X({\rm 7-30 keV}) = 3.2^{+2.7}_{-1.5} \times 10^{-3} 
S_{\gamma}({\rm 30-400 keV})^{0.643\pm0.046}$.  For the value of $S_{\gamma}$
we find for XRF~030723 from the power-law model, this yields the 
expectation that
$S_X=2.0^{+0.4}_{-0.6} \times 10^{-7}$ erg cm$^{-2}$.  This
is consistent with the observed value $S_X=1.6^{+0.3}_{-0.4} \times 10^{-7}$
erg cm$^{-2}$.  However, for the value of $S_{\gamma}$
we find from the cutoff power-law model
(which provides
a better fit to the XRF~030723 WXM/FREGATE data than does the power-law
model)
the \citet{barraud03} correlation
yields the expectation that
$S_X=3.0^{+2.0}_{-1.4} \times 10^{-8}$ erg cm$^{-2}$.  This
flux is a smaller than, and marginally inconsistent with, the observed value 
$S_X=1.6 \pm 0.5 \times 10^{-7}$ erg cm$^{-2}$.
In any case, XRF~030723 has a value for $S_X/S_{\gamma}$ near the
dividing line between XRFs and ``X-ray Rich'' GRBs ($S_X/S_{\gamma}=1$), 
and it is likely
an important ``bridge'' event between few keV XRFs like
XRF~020903 and classical GRBs and X-ray Rich GRBs.

\begin{figure}[ht]
\centering
\resizebox{30pc}{!}{\includegraphics{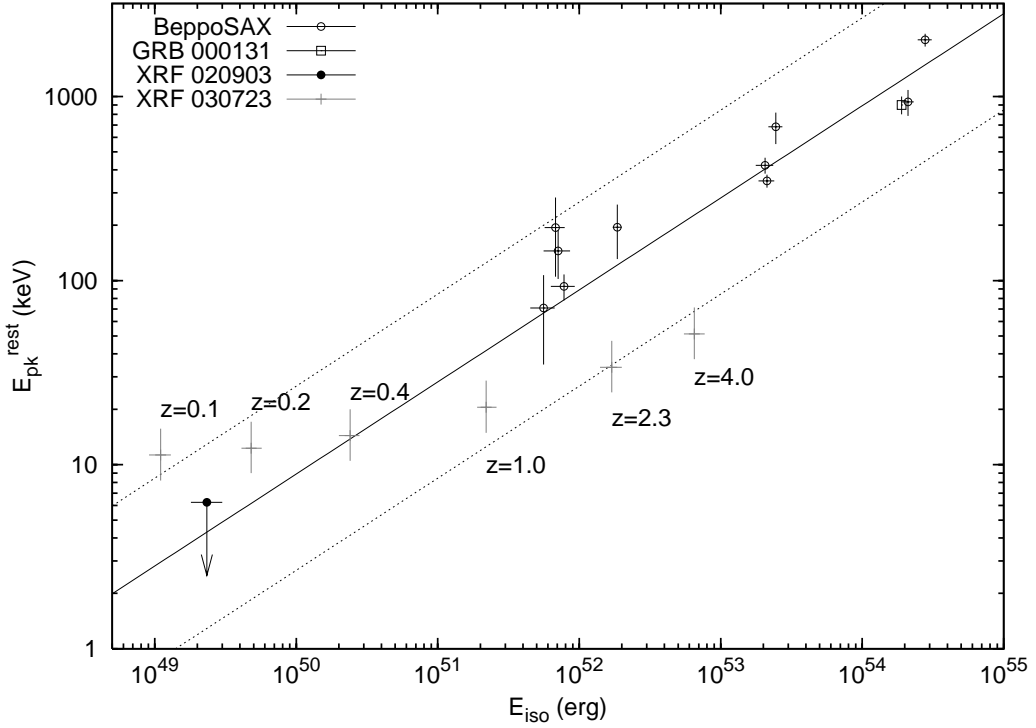}}
\caption{
\small
The trajectory with redshift of XRF~030723 through the $E_{\rm
pk}$-$E_{\rm iso}$ plane (crosses), where $E_{\rm iso}$
is the isotropic-equivalent radiated energy between 1-10$^{4}$ keV and
$E_{\rm pk}$ is the peak of the $\nu F_\nu$ spectrum, both measured
in the rest frame of the burst.  The solid line is given by the equation
 $E_{\rm pk} = 89 (E_{\rm iso}/10^{52}{\rm erg})^{0.5}$ keV.  The dotted
 region around this curve represents the 90\% confidence region, where
 we describe the intrinsic scatter in the \citet{amati02} relation as
 lognormal, with standard deviation 0.3.  The filled circle in the lower
 left-hand corner is the location of XRF~020903.  The ten open circles
 are the \beppo~GRBs reported by \citet{amati02}.  Given the constraint
 from the optical that $z<2.3$, the XRF~030723 data (like the XRF~020903
 data) are consistent with an extension of the \citet{amati02} relation to
 XRFs.
}
\label{fig:amati}
\end{figure}

\citet{sakamoto03}
find that XRF~020903 supports an extension of the \citet{amati02} 
correlation to XRFs.
Although no redshift has been measured for XRF~030723, we have a constraint
from the optical that $z<$2.3 (see Section \ref{sec:opt_aft}) and
possible indications from a rebrightening in the optical that the redshift 
may be lower yet.  The redshift is likely not much lower than $z=0.5$
due to the faintness of the host galaxy \citep{fynbo04a}.
In Figure \ref{fig:amati}, 
we plot the trajectory of XRF~030723 with redshift, determined from
fits of the cutoff power-law model, through the $E_{\rm
pk}$-$E_{\rm iso}$ plane.  For the
range of redshifts which agree with the optical constraints, the 
\hete~XRF~030723 data also support an extension of the \citet{amati02} 
relation to XRFs.
Taking this evidence as proof that XRF~030723 also satisfies the
\citet{amati02} relation,
we can turn the problem around and use the \citet{amati02} relation as 
prior information in our
spectral fitting.  This makes it possible to derive the best-fit redshift 
and 90\% confidence 
interval: $z=0.38^{+0.36}_{-0.18}$.  We also derive
$E_{\rm iso}= 2.1^{+8.7}_{-1.6} \times 10^{50}$ erg, and
$E_{\rm pk}=12.9^{+7.8}_{-4.4}$ keV.
We assume a lognormal distribution with standard deviation
0.3 for the intrinsic scatter in the \citet{amati02} relation.
We assume a cosmology with
(h,$\Omega_m$,$\Omega_{\Lambda}$)$=$(0.65,0.3,0.7)
Furthermore, we can motivate $z\lessim 1.0$,
using only the XRF data in the observer frame: we find
$\hat z=0.59$ for the \citet{atteia03} redshift indicator.  Use of the redshift
indicator $\hat z$ is justified in \citet{atteia03} via a correlation observed
for 17 GRBs detected by \hete~and \beppo.

The conclusion that XRF~030723 satisfies the \citet{amati02} relation is
most strongly dependent upon our determination at 98.5\% confidence
in Section \ref{sec:hete_spec} 
that the $E^{\rm obs}_{\rm pk}$ of the burst is well below 100 keV.  If this is not the 
case, then XRF~030723 may actually be a sub-energetic GRB like GRBs 980425 and
031203 \citep[see,][]{soder04b,woos04}.  These bursts apparently do not
lie on the \citet{amati04} correlation.   In the discussion below, we assume
that XRF~030723 does satisfy the \citet{amati04} relation.

\subsection{Testing the Off-axis Jet Model}

The prompt spectrum for XRF~030723 is likely not that of a typical GRB,
which has been redshifted to lower energies.  A redshift $z\sim 20$ would
be required to produce the low observed $E^{\rm obs}_{\rm pk}$ value.
This is clearly at odds with the redshift constraint just determined and
also with the redshift constraint derived from the optical data.

\begin{figure}[ht]
\centering
\resizebox{30pc}{!}{\includegraphics{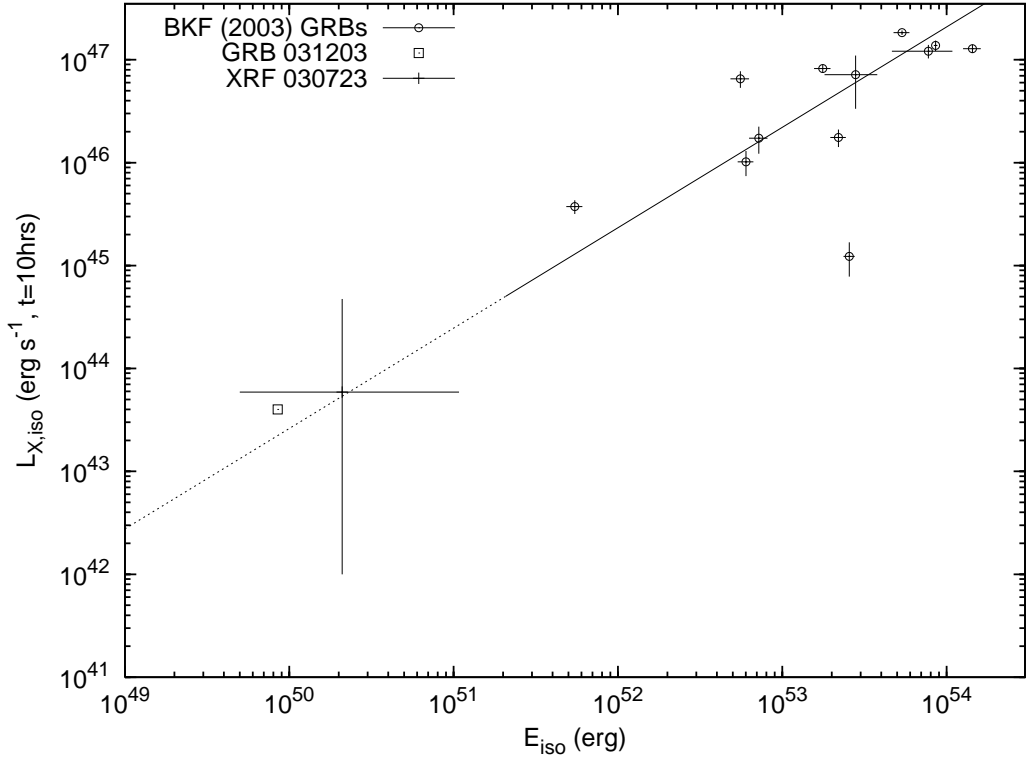}}
\caption{
\small
The location of XRF~030723 in the $L_{\rm X,iso}-E_{\rm iso}$ plane,
where $L_{\rm X,iso}$ is the isotropic-equivalent luminosity in the
2-10 keV band, adjusted to $t=10$ hrs after the burst, measured in
the rest frame of the burst.  For XRF~030723, we include the errors in 
the X-ray flux, temporal index ($\alpha=-1.0\pm 0.1$),
and redshift.  The XRF~030723 data extend a correlation
found by \citet{bkf} for GRBs.
$E_{\rm iso}$ values for the \citet{bkf} GRBs were taken from 
\citet{bloom03}.  Following \citet{kouv04}, we also plot the value for the
{\it INTEGRAL}~burst GRB~031203.
The solid line represent the best fit linear regression of \citet{bkf},
based on a flat-top, single-component jet model.  We extend this
line to jet opening angles $\theta_{\rm jet}>\pi$ (dotted region).
}
\label{fig:bkf}
\end{figure}

Using the 
redshift constraint derived assuming the \citet{amati02} relation,
we can compare the rest-frame X-ray afterglow and burst luminosities 
with values found for GRBs.  
It is interesting to note that
the $E_{\rm iso}$ we derive for the burst is approximately the
``standard energy'' found for GRBs by \citet{frail01}.  
This would appear to indicate that
XRF~030723 was an approximately spherical (i.e. non-jetted) explosion,
with an energy release typical of core-collapse supernovae.
However, the light curve break at $t\sim 1.5$ days (Section \ref{sec:opt_aft})
argues very strongly for jetted emission.  
Moreover, the afterglow energetics suggest further problems with this simple
picture.
Similar to the tight clustering of burst energies found by \citet{frail01},
\citet{bkf} find that
the isotropic equivalent emission in the X-ray afterglow 
(at 10 hours after the GRB)
also becomes tightly clustered if one accounts for differences in the
jet half-opening angles $\Delta \theta$.  As displayed
in Figure \ref{fig:bkf}, we find that XRF~030723 extends the \citet{bkf}
correlation to much lower luminosities.  In the case of XRF~030723, 
the simple 
flat-top jet picture would imply $\Delta \theta> \pi$, which violates
the model for both the burst and afterglow.  This appears to be the 
case for GRB~031203 
\citep[see,][]{watson04,sazo04} as well.  
The $E_{\rm iso}$ values for these
under-energetic bursts (including also XRF~020903) are less than the minimum 
energies permitted 
in the simple model where GRBs and XRFs have a standard energy 
release, with the observed energy determined by the width of the flat-top GRB 
jet.   This is also the case for GRB~980425, if the association with 
SN~1998bw is correct \citep{kouv04}.  

We can seek to retain a unified picture for GRBs and XRFs by
modifying the energetics of the explosion, as in the
dirty fireball \citep{dermer99,huang02} or clean fireball
\citep{moch03} models, and/or
by adopting a more complex description of the jet.
In Section \ref{sec:opt_aft} we summarize extensions of the
simple one component, on-axis, flat-top jet picture, which were found necessary 
to account for the complex behavior of the XRF~030723 optical 
afterglow.  In particular, several authors suggest the possibility 
of a jet viewed off-axis.  We can estimate
that $E^{\rm obs}_{\rm pk}$ would decrease by a factor
$\delta = 1+\gamma^2 (\theta_v - \Delta \theta)^2$
\citep[see, e.g.,][]{yama02}
if the jet with Lorentz factor
$\gamma$ were viewed at angle $\theta_v$ from its center.
A typical GRB would have $E^{\rm obs}_{\rm pk}\sim 200$ keV and 
$\gamma\sim 150$.
Our measured $E^{\rm obs}_{\rm pk}$ then implies $\theta_v - \Delta 
\theta \approx 1.8$ degrees.  The on-axis $E_{\rm iso}$ would scale up 
by factor $\sim \delta^{3}$ to $E_{\rm iso}\sim 2\times 10^{54}$ erg, which is
similar to values calculated for classical GRBs.  At early
times, a burst and afterglow viewed in such fashion would exhibit a
flat or possibly rising light curve due to the competing effects of the
fading afterglow flux and the jet expanding into the line of sight.  At
late times, the flux would break and decay as $t^{-2}$, which is the
canonical late time decay law observed for GRBs.  For a typical GRB
circumburst density of $1$ cm$^{-3}$, and using the formulas 
of \citet{sph99}, the break would occur 
at $t\sim 1.5$ days---as observed for
the optical light curve (and possibly X-ray light
curve; Section \ref{sec:rebright}) of XRF~030723---if $\Delta \theta \sim 4$ 
degrees.   Hence,
off-axis jets can explain the complex behavior of the optical
afterglow flux.  \citet{huang04} also argue for off-axis
viewing.  In their model, a narrow jet with parameters similar to those
above leads to the afterglow rebrightening at $t\gtrsim 9$ days
(see below and Section \ref{sec:opt_aft}).
The early-time afterglow light curve is due to
an additional wide ($\Delta \theta
\sim 17$ degree), baryon-loaded ($\gamma \sim 30$), and off-axis
($\theta_v \sim 21$ degrees) jet.

One important caveat is that the on-axis event would be a $\sim
3\sigma$ outlier on the \citet{amati02} plot.  Therefore, off-axis jetting
implies that XRFs should be sub-energetic outliers on the relation
\citep[see also, Figure 2 of][]{yama04}.  This may be the case for
XRF~030723 if the redshift is $\sim 0.1$ (Figure \ref{fig:amati}).
\citet{ghirl04} have recently discovered a correlation between
$E_{\gamma}$---the beaming corrected prompt energy 
release--and $E_{\rm pk}^{3/2}$,
which appears to be tighter than the $E_{\rm iso} \propto E_{\rm pk}^2$ 
correlation.
If we ignore the off-axis possibility and follow \citet{ghirl04} in 
calculating $E_{\gamma}=(1-\cos{\Delta \theta})\cdot E_{\rm iso}\sim
3\times 10^{48}$ erg, we likewise find that XRF~030723 satisfies the
correlation.  Here we use $\Delta \theta= 10$ degrees, because the observed
burst flux requires a broader jet in order to get the afterglow to break
at $t\sim 1.5$ days.  If we
consider the case where the off-axis GRB
discussed above is viewed on axis, we find $E_{\gamma} (\propto
E_{\rm iso}^{3/4}) \sim 5\times
10^{51}$ erg.  The on-axis event is a $\sim 4\sigma$ outlier.
Future prompt XRF observations will be critical to determine the
scatter in the correlations and to determine
whether or not XRFs demand a more complex jet model.

\subsection{X-ray Afterglow Rebrightening?}
\label{sec:rebright}

Given the apparent slow X-ray afterglow fade between our E1 and E2
\chandra~observations (possibly $\propto t^{-1}$, Section
\ref{sec:xray_fit}), the afterglow 
spectrum is consistent with a smooth extrapolation of the XRF
spectrum (in spectral shape and marginally in spectral flux (see Figure
\ref{fig:lc_opt})) from 
$t\sim 30$s.  These facts suggest a simple evolution,
where the X-ray afterglow has not undergone
a temporal break prior to $t\sim 13$ days.  A very similar argument
is made by \citet{amati04} for XRF~020427 based on data from \beppo.
Contrarily, we have evidence for complex temporal behavior
in the optical afterglow of XRF~030723, which was not available for
for XRF~020427.  It is possible
that the slow fade indicated by the ratio of our E1 and E2
\chandra~fluxes was produced by a more complex, time-varying
flux--possibly a rapid ($\propto t^{-2}$) fade followed by a
rebrightening ($\propto t^{3.5}$) as observed in the optical.

To test whether an X-ray rebrightening is or
is not present in the \chandra~data, we examine the arrival times
and energies of each source count in our E1 and E2 observations.  These
data are expected to provide further constraining information than would be
provided by the integrated fluxes alone.
We seek the model which maximizes the likelihood:
$$ {\cal L} (t_1,t_2,...,t_n,E_1,E_2,...,E_n) =
r(t_1,E_1)\cdot r(t_2,E_2)\cdot ...\cdot r(t_n,E_n)\cdot
\exp \left \{ -\int_{E_0}^{E_f} \int_{t_0}^{t_f} r(t',E') dt' 
dE' \right \}, $$
where the rate $r(t,E)$ is evaluated for each of $n$ observed
events, and where the time integral 
in the exponential is carried out for the good time
intervals of \chandra~data acquisition.   We
consider the data in the 0.5-8.0 keV band.
Assuming that the spectral slope does not evolve with time, we write the 
the event rate as $r(t,E)=c(E)*\phi(t)$, where $c(E)$ is the best 
fit count rate model from Section \ref{sec:xray_fit} 
(see also Figure \ref{fig:030723_sp}), and $\phi(t)$ is 
the time-dependent normalization.  For $\phi(t)$ we use a broken
power-law:
\begin{equation}
\label{EQUATION:broken_power_law}
\phi(t) = \left \{
        \begin{array}{lll}
   \phi_0 {\left(\frac{t}{t_b}\right)}^{\alpha_1}, & {\rm if} &  t \le t_b \\
   \phi_0 {\left(\frac{t}{t_b}\right)}^{\alpha_2}, & {\rm if} &  t > t_b,
        \end{array}
       \right.
\nonumber
\end{equation}
with $t_b=9$ days, as suggested by the optical decay curve.
The source data are extracted from 2.83\arcsec~and 0.84\arcsec~radius
regions, respectively, for E1 and E2.  This yields 76 and 74 events, 
respectively, for E1 and E2.  We ignore the effect of the $\sim 1$ 
background event in each epoch.

Fitting first for the case $\alpha_1=\alpha_2=\alpha$ (i.e. no break),
we find $\alpha=-0.9^{+0.1}_{-0.2}$, in agreement with the index determined
above.  Next, we allow the indices to vary independently.  We find
$\alpha_1=-2.1^{+2.0}_{-2.5}$ and $\alpha_2=3.5^{+8.7}_{-7.4}$.  
These
best-fit values are quite similar to their respective values
measured from the optical data.  Comparing
the likelihood calculated for each of these models, we have
$-2\Delta \log({\cal L}) = 1.08$.  This quantity is expected to be
approximately $\chi^2$ distributed with one degree of freedom.  Hence,
the rebrightening model is preferred by the data at approximately 70\% 
confidence.  We determine a similar estimate (68\% confidence that the
normalized event rate versus fraction of time observed is different in E1 
from that in E2) from a KS-test of the
photon arrival time distributions, ignoring the photon energies.  This
estimate does not require an event rate model.

\subsection{The Nature of the Rebrightening}

In Section \ref{sec:rebright}, we discuss marginal evidence 
that the X-ray afterglow flux tracks the optical afterglow flux.
We know of one case where a rebrightening in the X-ray afterglow
was tracked by the optical afterglow: GRB~970508, with a 
rebrightening at $t\sim 1$ day \citep{piro98}.
Because it also common for broad-band fades as $t^{-2}$ to occur
at late times, we regard a slow X-ray fade accompanying a considerably more
rapid optical fade after 1.5 days as unlikely.
If a slow fade were present in the optical
and not in the X-rays, we would expect to see an evolution in the
spectral flux between the optical and X-ray bands.
However, the afterglow spectrum in the optical \citep[see,][]{fynbo04a}
and in
the X-rays is consistent with a power-law spectrum with photon index
$\Gamma=2$, and this spectral slope does not appear to evolve (in
the X-rays or in the optical) between our E1 and E2 \chandra~observations.
If we fit the optical flux model (Figure \ref{fig:lc_opt}) to the
\chandra~count rates in E1 and E2, then the spectral slope 
(photon index) connecting the two bands is found to be 
$\Gamma=2.0\pm 0.1$, consistent with the slopes in the optical
and X-ray data considered alone.
Within the context of the standard fireball shock model, this 
broad-band slope suggests
a slow-cooling synchrotron spectrum where the cooling break
frequency is below the optical band \citep[see, e.g.,][]{spn98}.

Alternatively, it may be that the X-ray fade at $t\gtrsim 1.5$ days is in fact 
slower than
the optical fade due to the onset of an inverse-Compton (IC) emission
component \citep[see, e.g., Figure 3 of][]{pNk00}.  This would require
a fairly dense circumburst medium with $n\sim 100$ cm$^{-3}$.  
We would also expect to observe a flattening out of the E2 spectrum relative
to E1 by $\Delta\Gamma \ge 0.5$ \citep{sariNesin01}. 
However, such a flattening is excluded by the \chandra~data at
97.6\% confidence ($\Delta \chi^2 = 5.1$ for 1 additional degree of 
freedom).

If the same mechanism is responsible for the rebrightening at $t\gtrsim 9$ 
days in both the optical and in the X-ray bands, then an 
explanation in terms of a supernova bump is incorrect.
If the bump were produced by a dust echo,
we might expect to have observed a softening of the X-ray spectrum
in E2, as was observed for the possible dust echo in GRB~031203 by
\citet{vaugh04}.  Another possibility is that the afterglow shock interacts
with a density enhancement.  \citet{lazro02} find that the          
afterglow spectrum beyond the synchrotron cooling break is
insensitive to density enhancements.  However, the more complex
Wolf-Rayet wind models of \citet{ram01} are capable of producing
significant late time bumps (see their Figure 5).
\citet{ram01} also suggest the possibility that a dense shell of material
from a progenitor wind could generate a bump as a Compton echo of the
prompt emission (see their Figure 6).  Shortward of the hard X-rays, this
bump would have a spectrum mirroring that of the prompt emission, as we
seem to observe.  This scenario requires off-axis viewing for the bump to
be a substantial contributor to the afterglow flux.  Also in the off-axis
picture, the two-component jet model of \citet{huang04} would produce
a rise in the X-rays as well as in the optical.

In the context of a supernova bump, \citet{fynbo04a} explain the spectral
reddening at $t\approx$27 days (i.e. after the bump) as due to UV 
line-blanketing.
Alternatively, reddening could change as function of time if the progenitor
has a complex mass loss history, including a Wolf-Rayet phase and a dusty wind 
\citep[see, e.g.,][]{mir03,ram01}.

\section{Conclusions}

We have determined the temporal and spectral properties of XRF~030723,
as detected and measured by the WXM and FREGATE instruments on \hete.
The duration of the burst is similar to that common in long-duration
GRBs.  Moreover, the burst spectral properties are similar to those
of GRBs detected by FREGATE, except that the burst peak energy $E^{\rm obs}_{\rm pk}$
is $\sim$10 keV rather that $\sim$200 keV.  The low burst fluence appears
to be consistent with an extension of the \citet{amati02} relation to XRFs.
We have derived power-law spectral parameters to the X-ray afterglow data 
taken in two epochs with \chandra.  The photon index
we derive is a typical value for long-duration GRBs, further strengthening
the link between XRFs and long-duration GRBs.  We have argued that
the X-ray data, particularly when compared to the optical afterglow data,
favor a complex jet structure and off-axis viewing, possibly also a
complex circumburst medium due to a progenitor wind.  We wish to stress the
importance of multi-band observations, including X-ray observations,
to separate out and quantify these effects.  We encourage further late 
time ($t\gtrsim 1$ week) observations with \chandra~and/or\xmm~to help
determine the nature of possible rebrightenings in future afterglow 
light curves.

\acknowledgments

We thank Harvey Tananbaum for his generous allocation of Director's 
Discretion Time for the {\it Chandra}~observations.  This research was 
supported in part by NASA contract NASW-4690.

\clearpage

\begin{table}
\begin{center}
\caption{Temporal properties of XRF 030723.}
\vspace{5mm}
\begin{tabular}{ccccc}\hline\hline
Energy Band & Counts & Rate$_{\rm peak}$ & t$_{50}$     & t$_{90}$    \\
   (keV)    &  (in $\sim$35s) & (1/s) & (s)         &   (s)       \\\hline
    2 - 5   & 624 & 43.3  & 11.3$\pm$1.3  & 24.1$\pm$3.5 \\
    5 - 10  & 530 & 34.8  & 14.4$\pm$3.6  & 32.5$\pm$3.0 \\
    10 - 25 & 357 & 40.3  & 13.6$\pm$4.0  & 29.8$\pm$5.8 \\
    2 - 25 & 1430 & 81.2  & 12.0$\pm$1.3  & 28.3$\pm$2.5 \\\hline
\end{tabular}
\end{center}
{\small
Note.---The quoted errors correspond to $\pm 1\sigma$.
Burst counts and rates in the WXM are determined by subtracting 
a linear (in time)
background model (horizontal dotted lines in Figure \ref{fig:wxm_lcs}).
}
\label{table:duration}
\end{table}

\begin{deluxetable}{ccccccc}
\tabletypesize{\small}
\tablewidth{0pt}
\tablecaption{Results of fits to the WXM$+$FREGATE spectrum of XRF~030723.}
\startdata\hline\hline
Model 	& kT 	& $\alpha$ 	& $\beta$ 	& $E_{\rm peak}^{\rm obs}$ 	& Norm.	& $\chi^{2}_{\nu}$ (DOF) \\
        & (keV) & 		& 		& (keV) 			&  &\\\hline

blackbody & 2.0$_{-0.4}^{+0.5}$ & & & & $8.6\pm 1.5 \times 10^{-2}$ & 1.065 (82)\\

power-law &  & & $-2.0\pm 0.2$ 	& & $2.1\pm 0.3$ & 0.835 (82)\\

cutoff power-law & & $-$1.0 (fixed) &  & 9.3$_{-2.5}^{+3.6}$ & $0.8^{+0.3}_{-0.2}$ & 0.809 (82)\\

Band & 	& $-$1.0 (fixed) & $-3.4^{+1.0}_{-...^*}$	& $9.4^{+2.0}_{-1.6}$ & $8.0^{+1.3}_{1.2} \times 10^{-3}$ & 0.816 (81)\\

Band & & $-$1.4$\pm +0.4$ & $-$3.0$^{+0.8}_{-...^*}$ & $9.7^{+6.0}_{-2.5}$ & $2.0^{+7.4}_{-1.5} \times 10^{-3}$ & 0.816 (80)\\

\enddata

{Note.---The quoted errors correspond to the 90\% confidence region,
except for the Band models, where errors are 68\% confidence.  \newline
* The data do not allow for a determination of the lower limits for the Band 
model $\beta$ parameter.}
\label{table:spec_pars}
\end{deluxetable}

\begin{table}
\begin{center}
\caption{Peak photon number and energy fluxes (in 1 s) and fluences in
various energy bands for XRF~030723.}
\vspace{5mm}
\begin{tabular}{lccc}\hline\hline
          & 2-30 keV & 30-400 keV & 2-400 keV 
\\
\hline
Peak flux (ph cm$^{-2}$ s$^{-1}$) & $2.0^{+0.4}_{-0.3}$ & 
$0.1\pm 0.1$ & $2.1 \pm 0.4$
\\
Peak flux (10$^{-8}$ ergs cm$^{-2}$ s$^{-1}$) & $2.6\pm 0.5$ &
$0.9^{+1.8}_{-0.7}$ & $3.5^{+1.8}_{-1.1}$\\
Fluence (10$^{-7}$ ergs cm$^{-2}$) & $2.9\pm 0.5$ &
$0.2^{+0.3}_{-0.1}$ & $3.0^{+0.7}_{-0.6}$\\\hline
\end{tabular}
\end{center}
{\qquad \qquad \quad Note.---All of the quantities in this table are
derived assuming a cutoff power-law model with $\alpha=-1$ for the 
spectrum.  The quoted
errors correspond to the 90\% confidence region.}
\label{table:peak_fluxes}
\end{table}

\begin{table}[t]
\begin{center}
\caption{Characteristics of the
three \chandra~sources detected within the SXC error region.}
\vspace{5mm}
\begin{tabular}{rlrr}
\hline
\# &    Chandra Name   &    E1 Cts (bg) & E2 Cts (bg) \\
\hline
1 & CXOU J214924.4-274248  & 78.5 (1.5) & 75.6 (2.4) \\
3 & CXOU J214926.9-274146  & 19.9 (3.1) & 121.8 (4.2) \\
4 & CXOU J214928.7-274211  & 16.2 (3.8) & 98.1 (4.9) \\
\hline
\end{tabular}
\end{center}
{\small
Note.---We estimate a position uncertainty of 1.4\arcsec.
A steady source would have an E2 count rate $\sim 6$ times
greater than the E1 count rate.
}
\label{table:counts_030723}
\end{table}

\clearpage

\clearpage


\begin{thebibliography}{6}


\bibitem[Amati et al.(2002)]{amati02}
 Amati, L., et al. 2002, \aap, 390, 81
\bibitem[Amati et al.(2004)]{amati04}
 Amati, L., et al. 2004, \aap, 426, 415
\bibitem[Atteia(2003)]{atteia03}
 Atteia, J.-L. 2003, \aap, 407, L1
\bibitem[Band et al.(1993)]{band93}
 Band, D.~L., et al. 1993, \apj, 413, 281
\bibitem[Barraud et al.(2003)]{barraud03} 
 Barraud, C., et al. 2003, \aap, 400, 1021 
\bibitem[Bloom, Frail, \& Kulkarni(2003)]{bloom03}
 Bloom, J.~S., Frail, D.~A., \& Kulkarni, S.~R. 2003, \apj, 594, 674
\bibitem[Berger et al.(2003)]{berger03}
 Berger, E., et al. 2003, Nature, 426, 154B
\bibitem[Berger, Kulkarni, \& Frail(2003)]{bkf}
 Berger, E., Kulkarni, S.~R., \& Frail, D.~A. 2003, \apj, 590, 379
\bibitem[Bond(2003)]{bond03}
 Bond, H.~E. 2003, \gcn, 2329
\bibitem[Butler et al.(2003a)]{butler03a}
 Butler, N., et al. 2003a, \gcn, 2328
\bibitem[Butler et al.(2003b)]{butler03b}
 Butler, N., et al. 2003b, \gcn, 2347
\bibitem[Costa et al.(1999)]{costa99}
 Costa, E., et al. 1999, \aaps, 138, 425
\bibitem[Dado, Dar, \& De Rujula(2003)]{dado03}
 Dado, S., Dar, A., \& De Rujula, A. 2003, astro-ph/0309294
\bibitem[Dermer, Chiang, \& B\"ottcher(1999)]{dermer99} 
        Dermer, C. D., Chiang, J., \& B\"ottcher, M. 1999, \apj, 513, 656
\bibitem[Dickey \& Lockman (1990)]{dickey1990} 
        Dickey \& Lockman 1990 ARAA, 28, 215
\bibitem[Dullighan et al.(2003a)]{dull03a}
 Dullighan, A., et al. 2003a, \gcn 2326
\bibitem[Dullighan et al.(2003b)]{dull03b}
 Dullighan, A., et al. 2003b, \gcn, 2336
\bibitem[Fox et al.(2003)]{fox03}
 Fox, D.~B., et al. 2003, \gcn, 2323
\bibitem[Frail et al.(2001)]{frail01}
 Frail, D. A., et al. 2001, \apj, 562, L55
\bibitem[Fynbo et al.(2004a)]{fynbo04a}
 Fynbo, J.~P.~U., et al. 2004a, \apj, 609, 962
\bibitem[Fynbo et al.(2004b)]{fynbo04b}
 Fynbo, J.~P.~U., et al. 2004b, astro-ph/0402264
\bibitem[Ghirlanda, Ghisellini, \& Lazzati(2004)]{ghirl04}
 Ghirlanda, G., Ghisellini, G., \& Lazzati, D. 2004, astro-ph/0405602
\bibitem[Henden(2003)]{henden03}
 Henden, A. 2003, \gcn, 2343
\bibitem[Heise et al.(2000)]{heise2000}  
 Heise, J., et al. 2000, in
 Proc. 2nd Rome Workshop:  Gamma-ray Bursts in the Afterglow
 Era, eds. E. Costa, F. Frontera, J. Hjorth (Berlin:
 Springer-Verlag), 16
\bibitem[Huang, Dai, \& Lu (2002)]{huang02}
        Huang, Y. F., Dai, Z. G. \& Lu, T. 2002, MNRAS, 332, 735 
\bibitem[Huang et al.(2004)]{huang04}
 Huang, Y.~F., et al. 2004, \apj, 605, 300
\bibitem[Kawai et al.(2003)]{kawai03}
 Kawai, N., et al. 2003, \gcn, 2412
\bibitem[Kippen et al.(2002)]{kippen2002} 
 Kippen, R. M., et al. 2002, in Gamma-ray Bursts and
 Afterglow Astronomy, eds. G. R. Ricker and R. Vanderspek (New
 York: AIP), 244
\bibitem[Kouveliotou et al.(2004)]{kouv04}
 Kouveliotou, C., et al. 2004, \apj, 608, 872
\bibitem[Lazzati et al.(2002)]{lazro02}
 Lazzati, D., et al. 2002, \aap, 396, 5
\bibitem[Lewin, van Paradijs \& Taam(1993)]{lewin1993}
 Lewin, W. H. G., van Paradijs, J.\& Taam, R. E. 1993, Space
 Sci. Rev., 62, 223
\bibitem[Liang \& Dai(2004)]{liang04}
 Liang, E.~W., \& Dai, Z.~G. 2004, \apj, 608, L9
\bibitem[Mirabal et al.(2003)]{mir03}
 Mirabal, N., et al. 2003, \apj, 595, 935
\bibitem[Mochkovitch et al.(2003)]{moch03}
 Mochkovitch, R., et al. 2004, in Proc. 3rd Rome Workshop: Gamma-ray
 Bursts in the Afterglow Era, eds. M. Feroci, F. Frontera, N. Masetti,
 L. Piro (San Francisco: ASP), 381
\bibitem[Panaitescu \& Kumar(2000)]{pNk00}
 Panaitescu, A., \& Kumar, P. 2000, \apj, 543, 66
\bibitem[Piro et al.(1998)]{piro98}
 Piro, L., et al. 1998, \aap, 331, L41
\bibitem[Prigozhin et al.(2003)]{prig03}
 Prigozhin, G., et al. 2003, \gcn, 2313
\bibitem[Ramirez-Ruiz et al.(2001)]{ram01}
 Ramirez-Ruiz, E., et al. 2001, MNRAS, 327, 829
\bibitem[Rosati et al.(2002)]{rosati02}
 Rosati, P., et al. 2002, \apj, 566, 667
\bibitem[Rykoff et al.(2002)]{rykoff03}
 Rykoff, E.~S., et al. 2003, \apj, 601, 1013
\bibitem[Sakamoto et al.(2003)]{sakamoto03}
 Sakamoto, T., et al. 2003, \apj, 602, 875
\bibitem[Sari, Piran, \& Narayan(1998)]{spn98}
 Sari, R., Piran, T., \& Narayan, R. 1998, \apj, 497, L17
\bibitem[Sari, Piran, \& Halpern(1999)]{sph99}
 Sari, R., Piran, T., \& Halpern, J.~P. 1999, \apj, 519, L17
\bibitem[Sari \& Esin(2001)]{sariNesin01}
 Sari, R., \& Esin, A.~A. 2001, \apj, 548, 787
\bibitem[Sazanov, Lutovinov, \& Sunyaev(2004)]{sazo04}
 Sazonov, S. Yu., Lutovinov, A. A., \& Sunyaev, R., A. 2004, Nature, 430, 646
\bibitem[Schlegel et al.(1998)]{schleg98}
 Schlegel, D., et al. 1998, \apj, 500, 525
\bibitem[Shirasaki et al.(2000)]{shirasaki2000} 
 Shirasaki, Y., et al. 2000, SPIE, 4012, 166
\bibitem[Shirasaki et al.(2002)]{shirasaki2002} 
 Shirasaki, Y., et al. 2003, SPIE, 4851, 1310.
\bibitem[Smith et al.(2003)]{smith03}
 Smith, D.~A., et al. 2003, \gcn, 2338
\bibitem[Soderberg, Berger, \& Frail(2003)]{soder03}
 Soderberg, A.~M., Berger, E., \& Frail, D.~A. 2003, \gcn, 2330
\bibitem[Soderberg et al.(2004a)]{soder04a}
 Soderberg, A.~M., et al. 2004, \apj, 606, 994
\bibitem[Soderber et al.(2004b)]{soder04b}
 Soderberg, A.~M., et al. 2004, Nature, 430, 648
\bibitem[Strohmayer et al.(1998)]{strohmayer1998} 
 Strohmayer, T. E., et al.  1998, \apj,  500, 873
\bibitem[Tominaga et al.(2004)]{tom04}
 Tominaga, N., et al. 2004, \apj, 612, L105
\bibitem[Vaughan et al.(2004)]{vaugh04}
 Vaughan, S., et al. 2004, \apj, 603, L5
\bibitem[Watson et al.(2004)]{watson04}
 Watson, D., et al. 2004, astro-ph/0401225
\bibitem[Woosley(2004)]{woos04}
 Woosley, S. 2004, Nature, 430, 623
\bibitem[Yamazaki et al.(2002)]{yama02}
 Yamazaki, R., Ioka, K., \& Nakamura, T. 2002, \apj, 571, L31
\bibitem[Yamazaki, Kunihito, \& Nakamura(2004)]{yama04}
 Yamazaki, R., Kunihito, I., \& Nakamura, T. 2004, ApJ, 606, L33
\bibitem[Zhang \& M\'esz\'aros(2002)]{zhang2002}
 Zhang, B., \& M\'esz\'aros, P. 2002, \apj, 581, 1236

\end{thebibliography}
\end{document}